\documentstyle[preprint,aps]{revtex}
\newcommand{\intercal}{T}

\begin{document}
\draft



\title{
On the Solution of the Number-Projected Hartree-Fock-Bogoliubov Equations
}

\author{J.A. Sheikh$^{(1)}$, E. Lopes$^{(1)}$ and P. Ring$^{(1)}$}

\address{
$^{(1)}$Physik-Department, Technische Universit\"at M\"unchen,
D-85747 Garching bei M\"unchen, Germany
}

\maketitle

\begin{abstract}

The numerical solution of the recently formulated number-projected 
Hartree-Fock-Bogoliubov equations is studied in an exactly soluble 
cranked-deformed shell model Hamiltonian. It is found that the solution
of these number-projected equations involve similar numerical effort
as that of bare HFB. We consider that this
is a significant progress in the mean-field studies of the quantum
many-body systems. The results of the
projected calculations are shown to be in almost complete agreement with
the exact solutions of the model Hamiltonian. The phase
transition obtained in the HFB theory as a function of the rotational
frequency is shown to be smeared out with the projection.

\end{abstract}


The mean-field models with effective forces have been quite 
successful in describing the gross features of the quantum 
many-body systems. Although, the mean-field approaches are appropriate
for systems with a very large number of particles, they have also 
been quite useful to describe the properties of finite quantum systems,
for instance the atomic nucleus. The ground-state properties of atomic
nuclei have been well described using the Hartree-Fock (HF) and 
Hartree-Fock-Bogoliubov (HFB) mean-field approaches with various 
effective interactions \cite{rs80}. However, the mean-field 
application to a finite system suffers from a fundamental 
problem that it leads to sharp phase transitions. The phase
transition is an artefact of the mean-field approach and is
not observed in the experimental data. The phase transition
obtained is due to the neglect of the
quantal fluctuations, which become quite strong for finite systems.

There are various methods in the literature to consider the 
quantal fluctuations on the
mean-field solution for the finite system. One very powerful method
is through the restoration of the broken-symmetries by employing
the projection methods \cite{rs80}. In the present work, we shall
consider the restoration of the gauge-symmetry associated with the
particle-number. It is known in the HFB studies that one often
obtains a phase transition from the superfluid paired-phase
to the normal unpaired-phase. This phase transition is
due to the fluctuations in the particle-number, the HFB wavefunction
does not have a well defined particle-number. In most of the analysis, 
the particle-number
fluctuations are treated in an approximate way by employing 
the Lipkin-Nogami prescription \cite{mcd93,gbd94,ver96}. However, 
it has been shown that this
approach also breaks down at high-rotational frequencies and as a 
matter of fact violates the variational principle \cite{fo96}. 

The exact particle number-projection can be performed by using the 
gradient methods \cite{rs80}. But this approach is numerically 
quite involved and has been
applied to only separable interactions with restricted model
spaces \cite{ER.82a,ER.82b}.
There has been an unresolved issue whether HFB like equations 
can be obtained with the projected-
energy functional. This problem has been recently solved \cite{sr2000}
and it has been 
shown that it is possible to obtain
the HFB equations from an arbitrary real energy functional which is completely
expressible in terms of the density-matrix and the pairing-tensor. The
projected-energy functional can be expressed in terms of the density-matrix
and the pairing-tensor and one obtains the HFB equations with modified 
expressions for the pair-gap and the Hartree-Fock potential. The
expressions for these quantities acquire a relatively simple form for the
case of particle number-projection \cite{sr2000}.

To check the applicability of the number projected-HFB (PHFB) formalism, 
detailed numerical analysis is carried out in 
a simple cranked-deformed shell model Hamiltonian \cite{snrp89}.
Although, this model
cannot be used directly to study the experimental data, but it contains
all the basic ingredients of a more realistic model. The advantage 
in this model is that it can be
solved exactly and it is possible to check the accuracy of an approximate
method. We consider that it is quite instructive to test the 
number-projection method
in a cranking model as the Coriolis forces destroy the pair-correlations
and the results become quite sensitive to the treatment of the
pairing-interaction. 
As we shall demonstrate, the present projection method reproduces almost 
exactly the results of the shell model calculations for all the
cases studied.

The model Hamiltonian consists of a cranked deformed
one-body term, $h^\prime$ and a scalar two-body delta-interaction 
\cite{snrp89}. The one-body term is the familiar
cranked-Nilsson mean-field potential which takes into account of the
long-range part of the nucleon-nucleon interaction. The residual short-range
interaction is specified by the delta-interaction. 
The deformed shell model Hamiltonian employed is given by
\begin{eqnarray}
H&=&h^{\prime}_{def} + V_{2}, \label{E5010}\\
H&=&h_{def} - \omega J_x + V_{2},  \label{E5011}
\end{eqnarray}
where, 
\begin{equation}
h_{def}=-4\kappa \sqrt{\frac{4\pi }{5}}\sum_{m}<jm|Y_{20}|jm>c_{jm}^{\dagger
}c_{jm},  \label{E5021}
\end{equation}
and 
\begin{equation}
V_{2}={\frac{1}{2}}\sum_{LM}E_{L}^{{}}A_{LM}^{\dagger }A_{LM}^{{}},
\label{E5031}
\end{equation}
with $A_{LM}^{\dagger }=(c_{j}^{\dagger }c_{j}^{\dagger })_{LM}$ and $%
A_{LM}^{{}}=(A_{LM}^{\dagger })^{\dagger }$. For the
antisymmetric-normalized two-body matrix-element ( $E_{J}^{{}}$ ), we use
the delta-interaction which for a single j-shell is given by \cite{gb} 
\begin{equation}
E_{L}^{{}}=-G{\frac{(2j+1)^{2}}{2(2L+1)}}\left[ 
\begin{array}{ccc}
j & j & L \\ 
\frac{1}{2} & -\frac{1}{2} & 0
\end{array}
\right] ^{2}, 
\label{E5041}
\end{equation}
where the symbol $[~~]$ denotes the Clebsch-Gordon coefficient.
We use 
$G=g\int R^4_{nl}r^2 dr$ as our energy unit and the deformation
energy $\kappa$ is related to the deformation parameter $\beta$.
For the case of $h_{11/2}$ shell, $\kappa$=2.4 approximately
corresponds to $\beta=0.23$.

It has been shown in ref. \cite{sr2000} that the variation of the number
projected-energy functional results in the HFB matrix equation

\begin{equation}\label{hfb}
{\cal H}'\left( \begin{array}{c} U\\V \end{array}\right)=E'_i \left( 
\begin{array}{c} U\\V \end{array}\right),
\end{equation}
where
\begin{eqnarray}
{\cal H}'=&
\left( \begin{array}{cc} \varepsilon'_{n_1 n_2}+
\Gamma_{n_1 n_2}+\Lambda_{n_1 n_2}
-\lambda \delta_{n_1 n_2} & \Delta_{n_1 n_2} \\
-\Delta^{*}_{n_1 n_2}&-\varepsilon^{\prime *}_{n_1 n_2}-
\Gamma^*_{n_1 n_2}-\Lambda^*_{n_1 n_2}
+\lambda \delta_{n_1 n_2} 
\end{array}\right).\label{hfbh}&
\end{eqnarray}
The number-projected expressions for $\varepsilon'_{n_1 n_2}$,$\Gamma_{n_1 n_2}, \Lambda_{n_1 n_2}$ 
and $\Delta_{n_1 n_2}$ are given by

\begin{eqnarray}
\varepsilon' &=& 
\frac{1}{2}\int d\phi \,\,y(\phi )\left(Y(\phi)
{\rm Tr}[e'\rho(\phi)] +
[1-2ie^{-i\phi}\sin\phi\rho(\phi)]e'C(\phi)\right) \nonumber \\
&& ~+~h.c.
\label{E201}
\\
\Gamma &=&
\frac{1}{2}\int d\phi \,\,y(\phi )\left(Y(\phi)
\frac{1}{2}{\rm Tr}[\Gamma(\phi)\rho(\phi)] +\frac{1}{2}
[1-2ie^{-i\phi}\sin\phi\rho(\phi)]\Gamma(\phi)C(\phi)\right) \nonumber \\
&& ~+~h.c.
\label{E202}
\\
\Lambda &=&
-\frac{1}{2}\int d\phi \,\,y(\phi )\left(Y(\phi)
\frac{1}{2}{\rm Tr}[\Delta(\phi)\overline{\kappa}_{}^\ast(\phi)] -
2ie^{-i\phi}\sin\phi\; C(\phi)\Delta(\phi)\overline{\kappa}_{}^{\ast}\right)
\nonumber \\
&& ~+~h.c.
\label{E203}
\end{eqnarray}

\begin{equation}
\Delta =\frac{1}{2} \int d\phi\;y(\phi ) e^{-2i\phi }
C\left( \phi \right) \Delta
(\phi )-(..)^{\intercal } \label{E204},  
\end{equation}
with
\begin{eqnarray}
\Gamma _{n_{1}n_{3}}^{{}}(\phi ) 
&=&\sum_{n_2n_4}\overline{v}_{n_1n_2n_3n_4}^{}
\rho_{n_4n_2}^{}(\phi) \label{E501},\\
\Delta _{n_1n_2}^{}(\phi ) 
&=&\frac{1}{2}\sum_{n_3n_4}\overline{v}_{n_1n_2n_3n_4}^{}
\kappa_{n_3n_4}^{}(\phi) \label{E502},\\
\overline{\Delta}_{n_3n_4}^{\ast}(\phi ) 
&=&\frac{1}{2}\sum_{n_1n_2}\overline{\kappa}^{\ast}_{n_1n_2}(\phi)
\overline{v}_{n_{1}n_{2}n_{3}n_{4}}^{{}} \label{E503},
\end{eqnarray}

\begin{eqnarray}
\rho (\phi ) &=&C(\phi )\rho, \\
\kappa (\phi ) &=&C(\phi )\kappa =\kappa C_{{}}^{\intercal }(\phi ),\\
\overline{\kappa }(\phi ) &=&e^{2i\phi }\kappa C_{{}}^{\ast }(\phi
)=e^{2i\phi }C_{{}}^{\dagger }(\phi )\kappa,\\
C(\phi ) &=&e^{2i\phi }\left( 1+\rho (e^{2i\phi }-1)\right) ^{-1},\\
x(\phi )&=&\frac{1}{2\pi }\frac{e^{i\phi (N)} \det (e^{i\phi})}
{\sqrt{\det C(\phi )}},\\
y(\phi) &=&\frac{x(\phi)}{\int dg\,x(\phi)},\;\;\;\int dg\,y(\phi)=1,  
\label{E29}
\end{eqnarray}
and
\begin{equation}
Y(\phi )=\,ie^{-i\phi}\sin\phi\;C(\phi )-i
\int d\phi ^{\prime }y(\phi^\prime)e^{-i\phi^\prime}\sin\phi^\prime\;
C(\phi^\prime).
\end{equation}
The quantities $\rho$ and $\kappa$ in the above equations are the 
HFB density-matrix and the pairing-tensor.
$e'$ in (\ref{E201}) are the single-particle energies
of the cranked-deformed potential (\ref{E5010}) and 
$\overline{v}$ in (\ref{E501}-\ref{E503}) is the uncoupled 
antisymmetric matrix-element of the two-body delta-interaction (\ref{E5041}).

The term designated by $\Lambda$ in (\ref{E203}) does not appear 
in the ordinary HFB
formalism and it can be immediately shown that it vanishes for the gauge-angle
$\phi=0$. This term orginates from the variation of the pairing-energy with
respect to the density-matrix. In normal HFB, the pairing-energy depends only
on the pairing-tensor, but the PHFB pairing-energy also depends
on the density-matrix through the norm-overlap. Actually, in the general
the norm-overlap depends on both density-matrix and the 
pairing-tensor \cite{sr2000}.
But for the special case of number-projection, the term in the
overlap-matrix which depends on the pairing-tensor can be rewritten
in terms of the density-matrix by using the HFB relation 
$(\rho-\rho^2=\kappa \kappa^\dagger)$. Due to this transformation, the
expression for $\Delta$ in (\ref{E204}) has a very simple appearance and 
reduces to the familiar form in the canonical representation 
\cite{sr2000,dmp64}. 

The integration in (\ref{E201}-\ref{E204}) over the gauge-angle
has been performed using the 
Gauss-Chebyshev quadrature method \cite{hit79}. In this method,
the integration over the gauge-angle is replaced by a summation. 
It can be shown \cite{hit79} that the optimal number of 
mesh-points in the summation
which eliminates all the components having undesired particle
numbers is given by 
\begin{equation}
 M = { \rm max } \left( {1 \over 2} N, \Omega-{1 \over 2} \right) +1, 
\end{equation}
where N is the number of particles and $\Omega$ is the degeneracy of
the single-j shell. In the present study with N=6 and $\Omega=6$, the
optimal number of points required is $M=4$.

In the present analysis of a single-j shell, the basis in which the
HFB matrix is constructed are the magnetic sub-states of $j=11/2$ with 
$m=(11/2,9/2,........,-9/2,-11/2)$. The summation indices
$n_1, n_2, n_3$ and $n_4$ 
in the all the expressions given above run over these magnetic states.
In order to check the dependence of the HFB and PHFB results on the
pairing interaction, the calculations have been performed
with monopole (L=0), monopole plus quadrupole (L=0 and 2), and with 
full delta-interaction. The results of the HFB and PHFB will be 
compared with the exact results for the three pairing-interactions.

The results of the cranking calculations with monopole-pairing force 
are compared in Fig. 1. In the 
three-pannels of the figure, we compare the total-energy $(E_{tot})$,
the pairing-energy $(E_{pair})$
and the alignment $(<J_x>)$ which is the expectation value of the 
angular-momentum along the rotational x-axis, as a function of the 
rotational frequency. The expressions for the total energy is
given by

\begin{equation}
E_{tot}=\int d\phi \,y(\phi )\left( H_{sp}(\phi )+H_{ph}(\phi )+H_{pp}(\phi
)\right) ,  \label{108}
\end{equation}
where
\begin{eqnarray}
H_{sp}(\phi ) &=&{\rm Tr}\left( e\rho (\phi )\right) , \\
H_{ph}(\phi ) &=&\frac{1}{2}{\rm Tr}\left( \Gamma (\phi )\rho (\phi )\right),\\
H_{pp}(\phi ) &=&-\frac{1}{2}{\rm Tr}\left( \Delta (\phi )%
\overline{\kappa }_{{}}^{\ast }(\phi )\right).
\end{eqnarray}
The expressions for the pairing-energy and the alignment are given
by

\begin{eqnarray}
E_{pair}&=&\int d\phi H_{pp}(\phi ), \\
<J_x>   &=&\int d\phi {\rm Tr}\left( J_x\rho (\phi )\right).
\end{eqnarray}
It can be easily shown that for the gauge-angle, $\phi=0$, the
normal HFB expressions for these quantities are recovered.

It is apparent from the top pannel of Fig. 1 that the results of the 
exact shell model
and the PHFB are very similar, the two curves are almost
indistinguishable for all the frequency points. The results of HFB
on the other hand deviate considerably from the exact results 
at lower frequencies.
However, it can be seen from Fig. 2 that the HFB results converge towards
the exact results with increasing rotational frequency. The HFB energy 
before the
bandcrossing at $\hbar \omega=0.5$G is shifted from the exact energy by
a constant amount and can be improved by renormalising the strength
of the pairing-interaction. Therefore for the  total energy, the HFB 
approach is not a poor approximation. The actual problem in HFB lies 
in the analysis of the pairing-energy and the alignment which are 
shown in the two lower-pannels of Fig. 1. The HFB pairing-energy 
has a finite value till
$\hbar \omega=0.45$G and then suddenly goes to zero at $\hbar \omega=0.5$G.
This transition is an artefact of the HFB approach as is clearly evident
from Fig. 1. The PHFB pairing-energy does drop at the bandcrossing but
has a finite value at all the rotational frequencies. This feature of the 
pairing- energy is also reflected in the alignment, which is
determined by the competition between the Coriolis forces and the pairing-
correlations. The alignment till $\hbar \omega=0.45$G is very similar in
all the three cases but then HFB value suddenly jumps to $<J_x>=10$, the
exact and the PHFB values on the on other hand do not show this sharp
transition. The gain in alignment at the first-crossing,
referred to as the (AB)-crossing, is quite similar in all the cases, however,
in HFB the crossing occurs somewhat earlier.

The results of pairing-interaction with monopole plus quadrupole terms are
shown in Fig. 2. The comparison among HFB, PHFB and the exact results is quite
similar to Fig. 2. The total HFB energy is shifted by a constant factor from
exact and PHFB energies before the bandcrossing. After the crossing, the
HFB results become closer to the exact. The HFB pairing-energy depicts a
transition and PHFB pairing-energy on the other hand
drops smoothly at the bandcrossing. However,
the HFB gain in alignment at the (AB)-crossing is much lower than 10
and has clearly a wrong behaviour after the crossing. 

The results with the full delta-interaction are presented in Fig. 3. The
HFB total-energy is now in a better agreement with exact and PHFB as compared
to the results shown in Figs. 1 and 2. In fact, it is evident by comparing
the three figures that the total-energy improves by including higher
multipoles in the pairing-interaction, maximum deviation is noted for
the monopole case. This appears to be in contradiction to our basic 
understanding of
the mean-field approach, one normally expects that HFB or BCS is a 
better approximation for the pure monopole-pairing. However, 
it should be mentioned that in our HFB and PHFB analysis, we have included
all the terms in the Hamiltonian. In particular, the
particle-hole contribution ($\Gamma$) amounts to about 6G in 
the total-energy and is
maximum with the full delta-interaction. If one excludes this contribution,
the discrepancy would be largest for the delta-interaction.

The pairing-energy in Fig. 3 again depicts a phase transition 
at $\hbar \omega=0.55$G
which is slightly higher than with monopole-interaction. The HFB (AB)-crossing
with full delta-interaction is now close to the exact and PHFB. The gain in
alignment at the (AB)-crossing is lower than 10 as in the case of
monopole and quadrupole pairing force. The overall agreement with full
delta-interaction appears to be better for HFB. The good agreement between
PHFB and exact on the other hand is independent of the interaction
used. 

To conclude, in the present work the number-projected HFB approach 
recently developed has been applied to an exactly soluble 
cranked-deformed shell Hamiltonian. 
The main motivation has been to check the numerical applicability of the
projection method. It is clear from the present study that the
projected Hartree-Fock-Bogoliubov approach gives an accurate description
of the yrast-states of the model Hamiltonian. The transition
from a superfluid to the normal phase obtained in the HFB theory is shown
to be smeared out with the projection.

We would like to stress that the main advantage of the present projection
method is that it has the same structure as that of normal HFB equations.  
Therefore, one can use the existing HFB computer codes and only the
expressions for the Hartree-Fock potential and the pairing-field need
to be redefined. Instead of the normal HFB fields, in the projection method
one needs to calculate the projected quantities as given by 
(\ref{E201}-\ref{E204}). In the present model study, we find that the 
numerical work involved
in the projection is similar to performing the bare HFB calculations.
For each rotational frequency, the average CPU time on Pentium (166MHz)
was 6.14s with projection as compared to 5.97s for normal HFB.
The present projection method, therefore, preserves
all the mathematical and computational simplicity of the HFB 
mean-field approach.


\begin{figure}
\caption{
The results of the total energy $(E_{tot})$, the 
pair-energy 
$(E_{pair})$ and the alignment $(J_x)$ for six-particles 
in a deformed $j=11/2$ orbitial using the monopole-interaction.
The PHFB results are indistinguisable from the exact shell models
results.}
\label{figure.1}
\end{figure}

\begin{figure}
\caption{
The results of the total energy $(E_{tot})$, the 
pair-energy 
$(E_{pair})$ and the alignment $(J_x)$ for six-particles 
in a deformed $j=11/2$ orbitial using the monopole plus quadrupole
interaction.
}
\label{figure.2}
\end{figure}

\begin{figure}
\caption{
The results of the total energy $(E_{tot})$, the 
pair-energy 
$(E_{pair})$ and the alignment $(J_x)$ for six-particles 
in a deformed $j=11/2$ orbitial using the full delta-interaction.
}
\label{figure.3}
\end{figure}

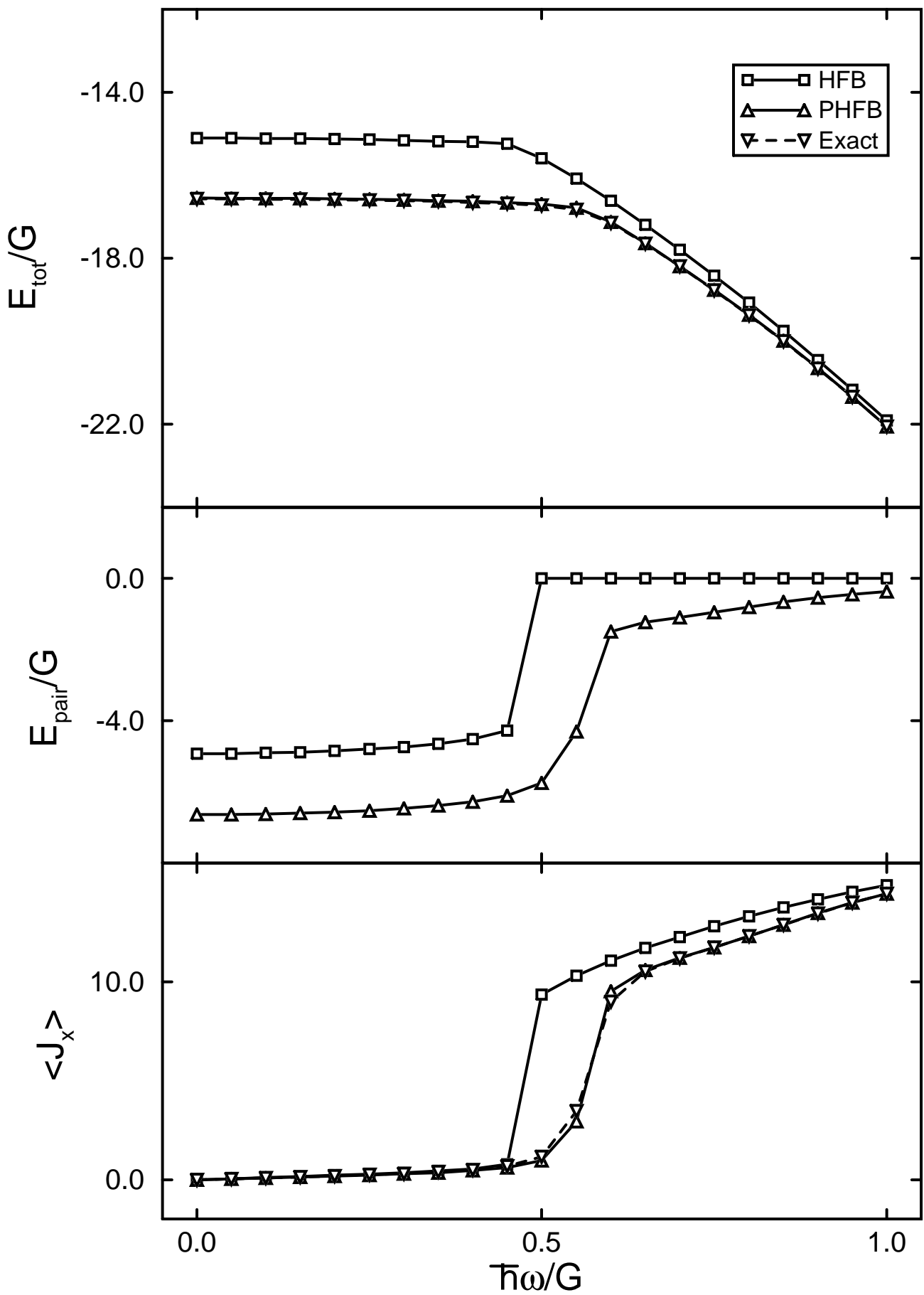
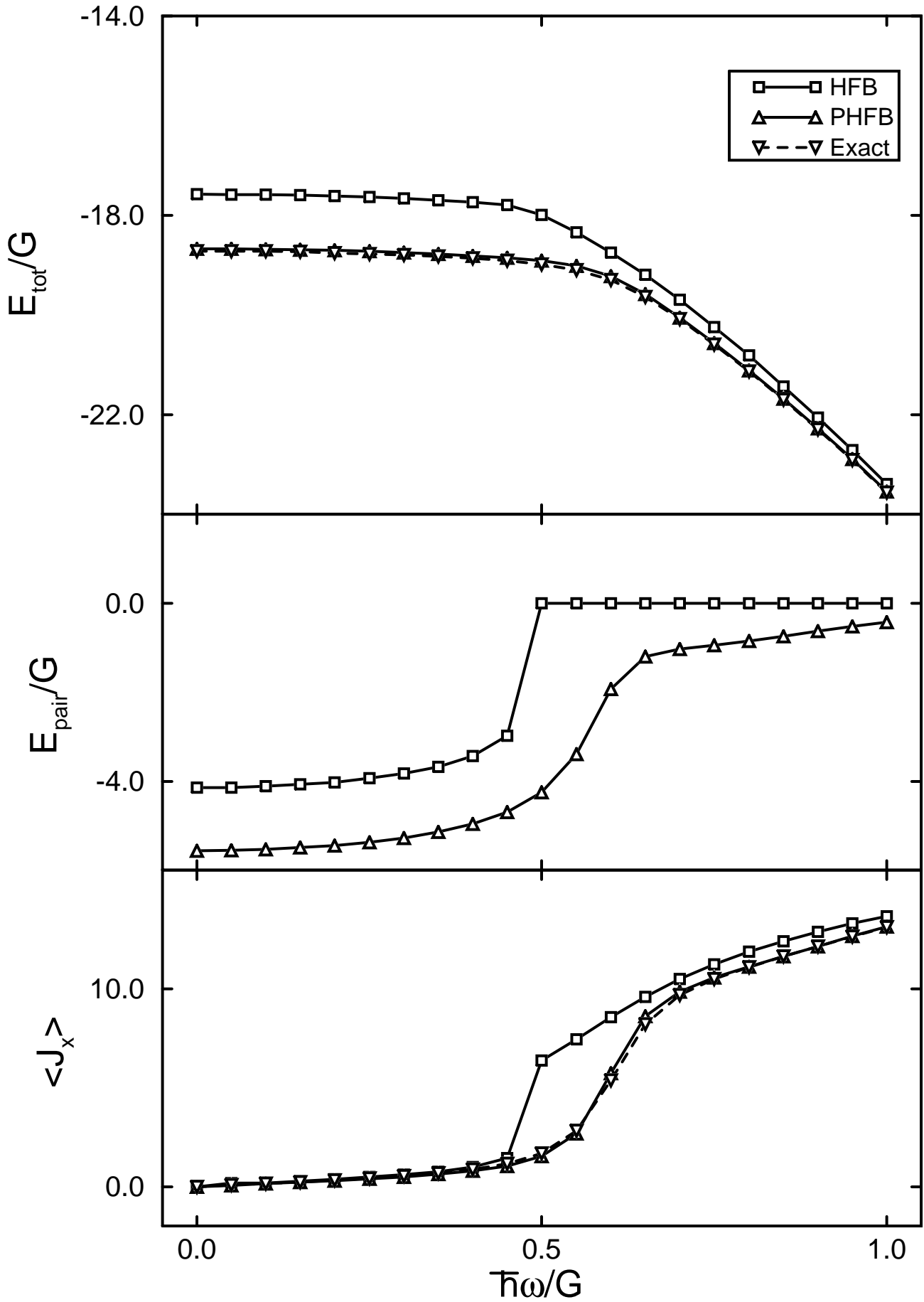
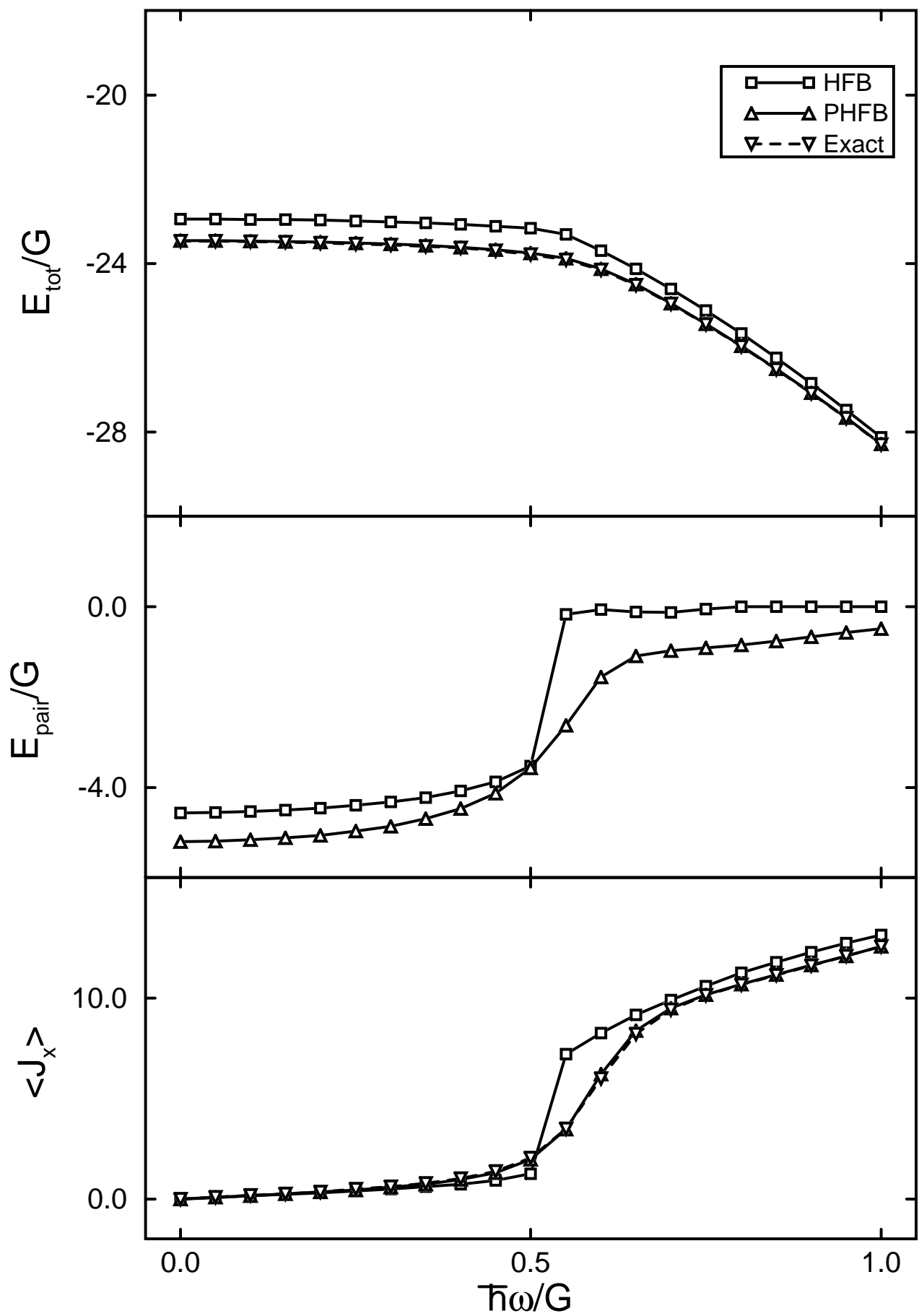
\end{document}